\documentclass[aps,prl,superscriptaddress,amssymb,reprint,nobibnotes]{revtex4-1}
\pdfoutput=1
\usepackage{amsmath,amssymb,latexsym}
\usepackage{amsthm}
\usepackage{amsfonts}
\usepackage{bbold}
\usepackage{graphicx}   
\usepackage{verbatim}   
\usepackage{color}      
\usepackage{subfigure}  
\usepackage{hyperref}   
\usepackage{natbib}
\usepackage[USenglish]{babel}

\newcommand{\bra}[1]{\langle #1|}
\newcommand{\ket}[1]{|#1\rangle}

\newcommand{\mszero}{m_\mathrm s = 0}
\newcommand{\mspmone}{m_\mathrm s = \pm1}
\newcommand{\msmone}{m_\mathrm s = -1}

\newcommand{\mizero}{m_\mathrm I = 0}
\newcommand{\mimone}{m_\mathrm I = -1}
\newcommand{\mipone}{m_\mathrm I = +1}
\newcommand{\be}{\begin{equation}}
\newcommand{\ee}{\end{equation}}

\newcommand{\cthirteen}{$^{13}$C}
\newcommand{\nfourteen}{$^{14}$N}

\newcommand{\ii}{\mathrm i}
\newcommand{\expe}{\mathrm e}
\newcommand{\textmicro}{$\mu$}

\begin{document}

\title{Unconditional quantum teleportation between distant solid-state qubits}
\author{Wolfgang Pfaff}
\altaffiliation[Present address: ]{Department of Applied Physics, Yale University, New Haven, CT 06511, USA}
\author{Bas Hensen}
\author{Hannes Bernien}
\author{Suzanne B.\ van Dam}
\author{Machiel S.\ Blok}
\author{Tim H.\ Taminiau}
\author{Marijn J.\ Tiggelman}
\author{Raymond N.\ Schouten}
\affiliation{Kavli Institute of Nanoscience Delft, Delft University of Technology, P.O. Box 5046, 2600 GA Delft, The Netherlands}
\author{Matthew Markham}
\author{Daniel J.\ Twitchen}
\affiliation{Element Six, Ltd., Kings Ride Park, Ascot, Berkshire SL5 8BP, United Kingdom}
\author{Ronald Hanson}
\email{r.hanson@tudelft.nl}
\affiliation{Kavli Institute of Nanoscience Delft, Delft University of Technology, P.O. Box 5046, 2600 GA Delft, The Netherlands}

\begin{abstract}
Realizing robust quantum information transfer between long-lived qubit registers is a key challenge for quantum information science and technology. Here we demonstrate unconditional teleportation of arbitrary quantum states between diamond spin qubits separated by 3 meters. We prepare the teleporter through photon-mediated heralded entanglement between two distant electron spins and subsequently encode the source qubit in a single nuclear spin. By realizing a fully deterministic Bell-state measurement combined with real-time feed-forward we achieve teleportation in each attempt while obtaining an average state fidelity exceeding the classical limit. These results establish diamond spin qubits as a prime candidate for the realization of quantum networks for quantum communication and network-based quantum computing.
\end{abstract}

\maketitle

The reliable transmission of quantum states between remote locations is a major open challenge in quantum science today. Quantum state transfer between nodes containing long-lived qubits \cite{Awschalom:2013in,2013Sci...339.1169D,2013Sci...339.1164M} can extend quantum key distribution to long distances \cite{1998PhRvL..81.5932B}, enable blind quantum computing in the cloud \cite{2012Sci...335..303B} and serve as a critical primitive for a future quantum network \cite{2008Natur.453.1023K}. When provided with a single copy of an unknown quantum state, directly sending the state in a carrier such as a photon is unreliable due to inevitable losses. Creating and sending several copies of the state to counteract such transmission losses is impossible by the no-cloning theorem \cite{1982Natur.299..802W}. Nevertheless, quantum information can be faithfully transmitted over arbitrary distances through quantum teleportation provided the network parties (named ``Alice'' and ``Bob'') have previously established a shared entangled state and can communicate classically \cite{1993PhRvL..70.1895B,Bouwmeester:1997wk,1998Sci...282..706F,Takeda:2013hn}.

\begin{figure*}
    \includegraphics{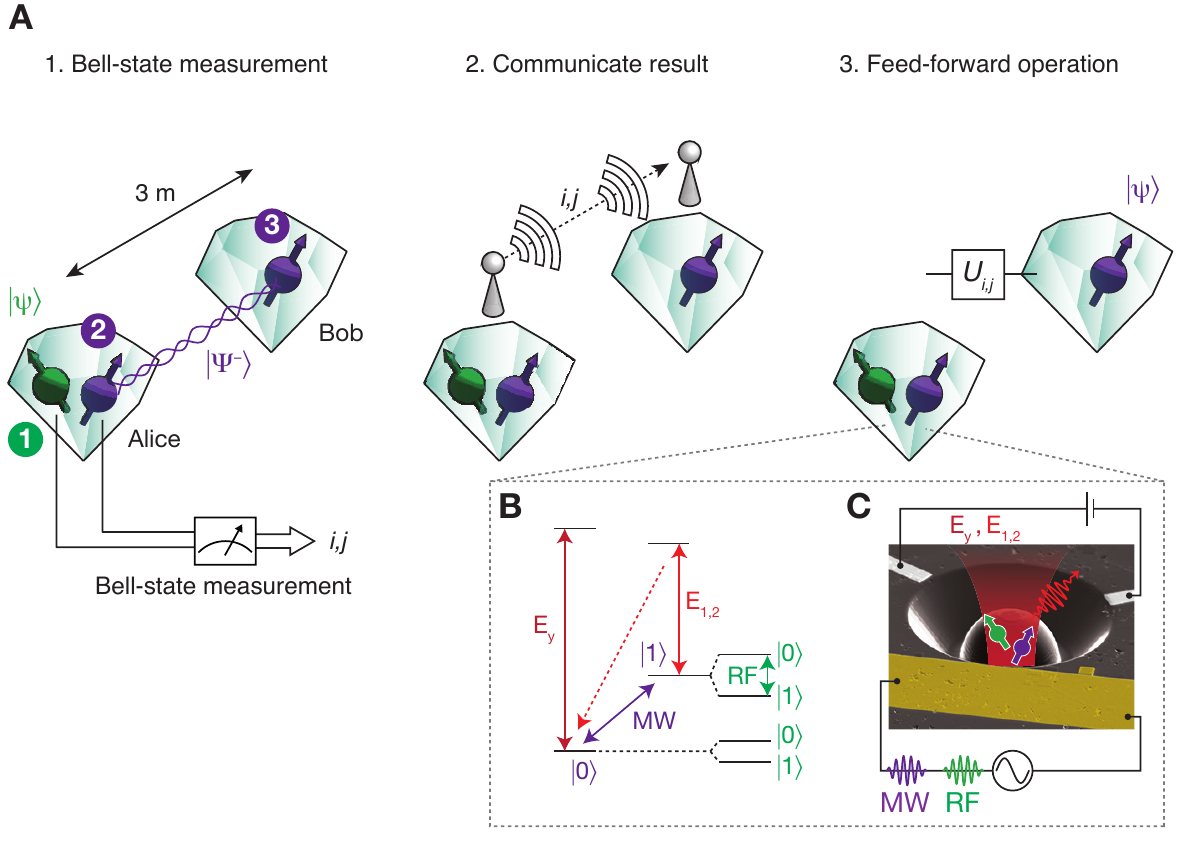}
    \caption{
    \label{fig1} 
    Teleportation scheme and system description. 
    \textbf{(A)} General scheme for teleportation. In our experiment Alice and Bob each control a single NV center in a single-crystal CVD-grown diamond by operating an independent cryogenic confocal microscope setup (T = 8\,K for Alice and T  = 4\,K for Bob). 
    \textbf{(B)} Energy level scheme and qubit control methods. The source state is encoded in Alice's nitrogen-14 spin (green) with basis states $\ket{0} \equiv \mizero$, $\ket{1} \equiv \mimone$. Two distant NV electronic spins (purple), with basis states encoded as $\ket{0} \equiv \mszero$ and $\ket{1} \equiv \msmone$, form the remote entangled pair shared by Alice and Bob. The electron spin is initialized by optical spin pumping on the NV center's $E_{1,2}$ transitions (bright red arrows), and read out by spin-selective optical excitation via the $E_y$ transition (dark red arrow)~\cite{Robledo:2011fs}. Microwave (MW) pulses allow for electron spin manipulation, and RF pulses are used to manipulate the nuclear spin when the electron is in state $\ket1$. 
    \textbf{(C)} Scanning electron microscope image of a diamond device, featuring a solid-immersion lens for enhanced collection efficiency, a stripline for spin manipulation by magnetic resonance, and electrodes for bringing the optical transitions of Alice and Bob on resonance using the d.c.\ Stark effect. 
    }
\end{figure*}

The teleportation protocol is sketched in Fig.~\ref{fig1}A. At the start, Alice is in possession of the state to be teleported (qubit 1) which is most generally given by $\ket\psi = \alpha\ket0 + \beta\ket1$. Alice and Bob each have one qubit of an entangled pair (qubits 2 and 3) in the joint state $\ket{\Psi^-}_{23} = (\ket{01}_{23} - \ket{10}_{23})/\sqrt2$. The combined state of all three qubits can be rewritten as
\begin{align}
    \ket\psi_1 \otimes \ket{\Psi^-}_{23} = \frac{1}{2} \big( 
        & \ket{\Phi^+}_{12} \otimes (\alpha\ket1_3 - \beta\ket0_3) \nonumber\\
        + & \ket{\Phi^-}_{12} \otimes (\alpha\ket1_3 + \beta\ket0_3) \nonumber\\
        + & \ket{\Psi^+}_{12} \otimes (- \alpha\ket0_3 + \beta\ket1_3) \nonumber\\
        - & \ket{\Psi^-}_{12} \otimes (\alpha\ket0_3 + \beta\ket1_3)
        \big),
\end{align}
where $\ket{\Phi^\pm} = (\ket{00}\pm\ket{11})/\sqrt2$ and $\ket{\Psi^\pm} = (\ket{01}\pm\ket{10})/\sqrt2$ are the four Bell states. To teleport the quantum state Alice performs a joint measurement on her qubits (qubits 1 and 2) in the Bell basis, projecting Bob's qubit into a state that is equal to $\ket\psi$ up to a unitary operation that depends on the outcome of Alice's measurement. Alice sends the outcome via a classical communication channel to Bob, who can then recover the original state by applying the corresponding local transformation.

Because the source qubit state always disappears on Alice's side, it is irrevocably lost whenever the protocol fails. Therefore, to ensure that each qubit state inserted into the teleporter unconditionally re-appears on Bob's side, Alice must be able to distinguish between all four Bell states in a single shot and Bob has to preserve the coherence of the target qubit during the communication of the outcome and the final conditional transformation. Several pioneering experiments have explored teleportation between remote nodes~\cite{Olmschenk:2009dr,Nolleke:2013ha,2013NatPh...9..400K} but unconditional teleportation between long-lived qubits~\cite{Awschalom:2013in,2013Sci...339.1169D,2013Sci...339.1164M} has so far only been demonstrated within a local qubit register~\cite{Riebe:2004cz,Barrett:2004kf,2013Natur.500..319S}.

Here we demonstrate unconditional teleportation between diamond spin qubits residing in independent setups separated by 3 meters. We achieve this result by fully separating the generation of remote entanglement (the preparation of the teleporter) from the two-qubit Bell-state measurement and feed-forward (the actual teleportation action). In particular, a photonic channel is used to generate heralded remote entanglement between two nitrogen-vacancy (NV) center electronic spins, while the teleportation protocol solely exploits matter qubits that – unlike photonic qubits – allow for a deterministic Bell-state measurement with current technology. The source state is encoded in a nuclear spin close to one of the NV electron spins after preparation of the teleporter. We preserve the target qubit's coherence by dynamical decoupling while the measurement outcome is forwarded and the final correction pulse is applied. This protocol ensures that the source state is successfully teleported in each of the experimental runs.

In our experiment Alice and Bob each operate an independent low-temperature confocal microscope setup that addresses a single NV center. The two NV electronic spins (labeled as qubits 2 and 3) are used as the distributed entangled pair that is the medium for teleportation. These spins can be initialized and read out in a single shot by spin-resolved optical excitation~\cite{Robledo:2011fs} and coherently manipulated using microwave (MW) pulses~\cite{2010Sci...330...60D} (Fig.~\ref{fig1}B). 

\begin{figure*}
    \includegraphics{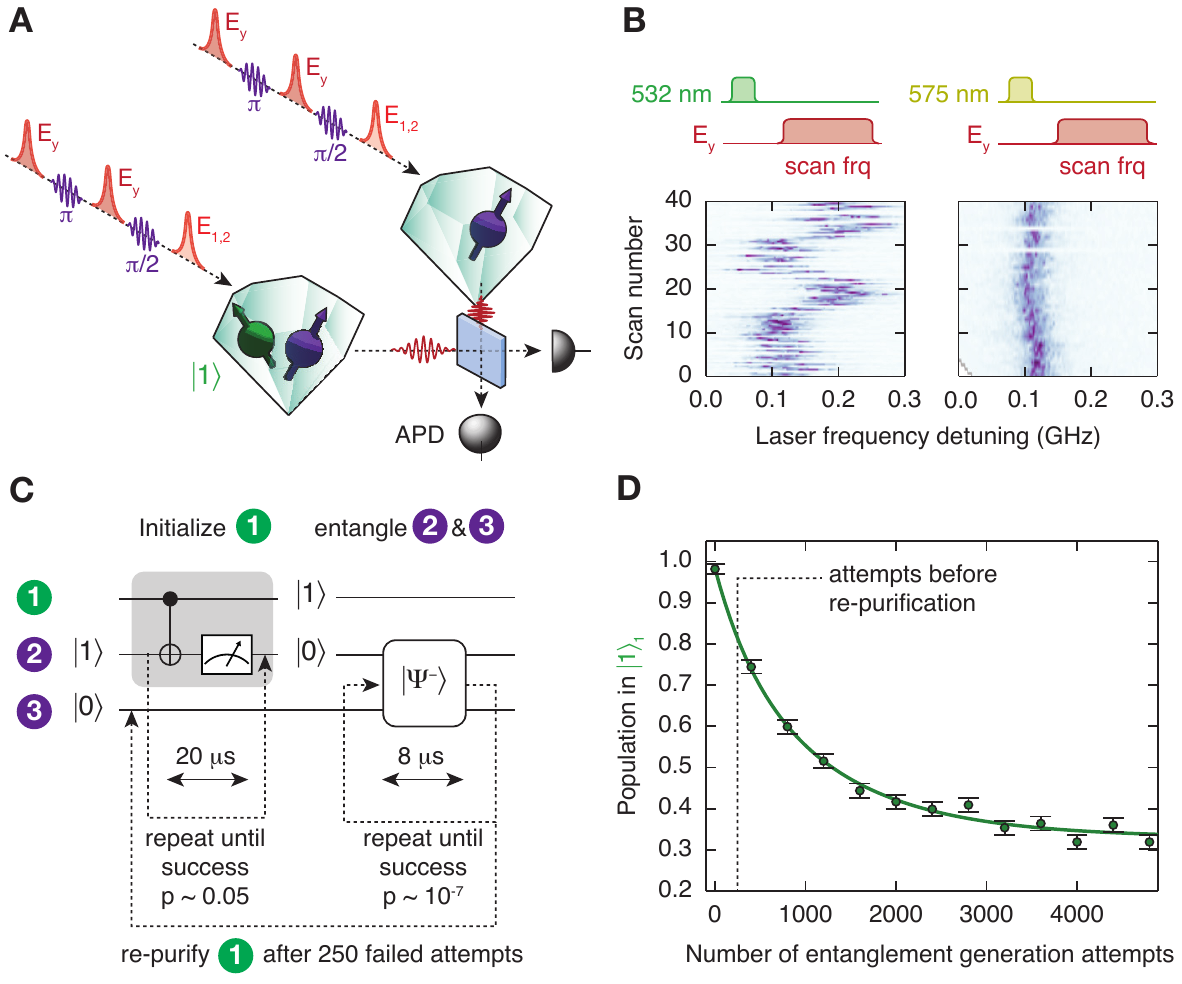}
    \caption{
    \label{fig2} 
    Preparation of the teleporter.
    \textbf{(A)} Schematic showing generation of remote entanglement. After initializing qubit 1 in $\ket1$ the following sequence is applied. First both qubit 2 and 3 are initialised in $\ket0$ by optical pumping. Then a combination of spin rotations and spin-selective optical excitation on $E_y$ creates local entanglement between spin and photon number at each node, followed by two-photon quantum interference and photon detection (for projecting qubits 2 and 3 onto an entangled state) using avalanche photo detectors (APDs) \cite{2005PhRvA..71f0310B,2013Natur.497...86B}. This sequence is repeated until successful. In the experiment the photons are guided through fibers to the beamsplitter and the APDs. 
    \textbf{(B)} Measurement of the frequency stability of the optical transition labeled $E_y$. We repeatedly apply a charge repump pulse and then scan a red laser (5\,nW) over the $E_y$ resonance. Spectral diffusion is strongly slowed down for the charge repump laser (50\,nW) on resonance with the NV$^0$ zero-phonon line at 575\,nm (right) compared to conventional off-resonant charge repumping using laser light (150\,\textmicro W) at 532\,nm (left) \cite{Siyushev:2013en}. For the scans using a 575\,nm repump laser we apply a strong laser pulse on resonance with NV$^-$ (50\,nW) before each scan to enforce ionization to NV$^0$. The red laser frequency shown is relative to 470.4636\,THz. Color encodes the photon count rate during the scan, darker indicates higher intensity. 
    \textbf{(C)} Circuit diagram for the periodic measurement-based re-initialization of the nuclear spin (qubit 1) in between remote entanglement generation attempts. Both the probability for success per attempt and the time duration of a single attempt are indicated for the initialization by measurement of qubit 1 and the generation of entanglement between qubits 2 and 3. 
    \textbf{(D)} Measured probability P($\ket1$) to preserve the initialized nuclear spin state $\ket1$ as a function of number of entanglement generation attempts $N_\text{ent}$. A fit (solid line) to a rate-equation model yields a probability of $(0.85 \pm 0.05) \times 10^{-3}$ per entanglement generation attempt that the nuclear spin flips. The dashed line marks the maximum number of attempts before the nuclear spin is re-initialized ($N_\text{ent} = 250$). }
\end{figure*}

To prepare the teleporter we initialize the electrons in the non-local entangled state $\ket{\Psi^-}_{23} = (\ket{01}_{23} - \ket{10}_{23})/\sqrt2$ through a recently demonstrated protocol~\cite{2005PhRvA..71f0310B,2013Natur.497...86B} that is based on local entanglement between electron spin and photon number and subsequent joint measurement of the photons (Fig.~\ref{fig2}A). Because successful entanglement generation is heralded by photon detection events, the protocol is robust against photon loss. Compared to the initial demonstration of this entangling protocol~\cite{2013Natur.497...86B} we have further enhanced the efficiency of photon collection from our device through an anti-reflection coating. Also, we have significantly improved both the spectral stability of the NV center's optical transition and the charge state initialization by resonant re-pumping on the neutral-charge state zero-phonon line~\cite{Siyushev:2013en} (Fig.~\ref{fig2}B). As a result we were able to increase the generation rate of the entangled state $\ket{\Psi^-}_{23}$ fivefold to $1/250\, \mathrm s^{-1}$ and improve the entangled state fidelity from 0.73 to an estimated 0.87 (see below).

\begin{figure*}
    \includegraphics{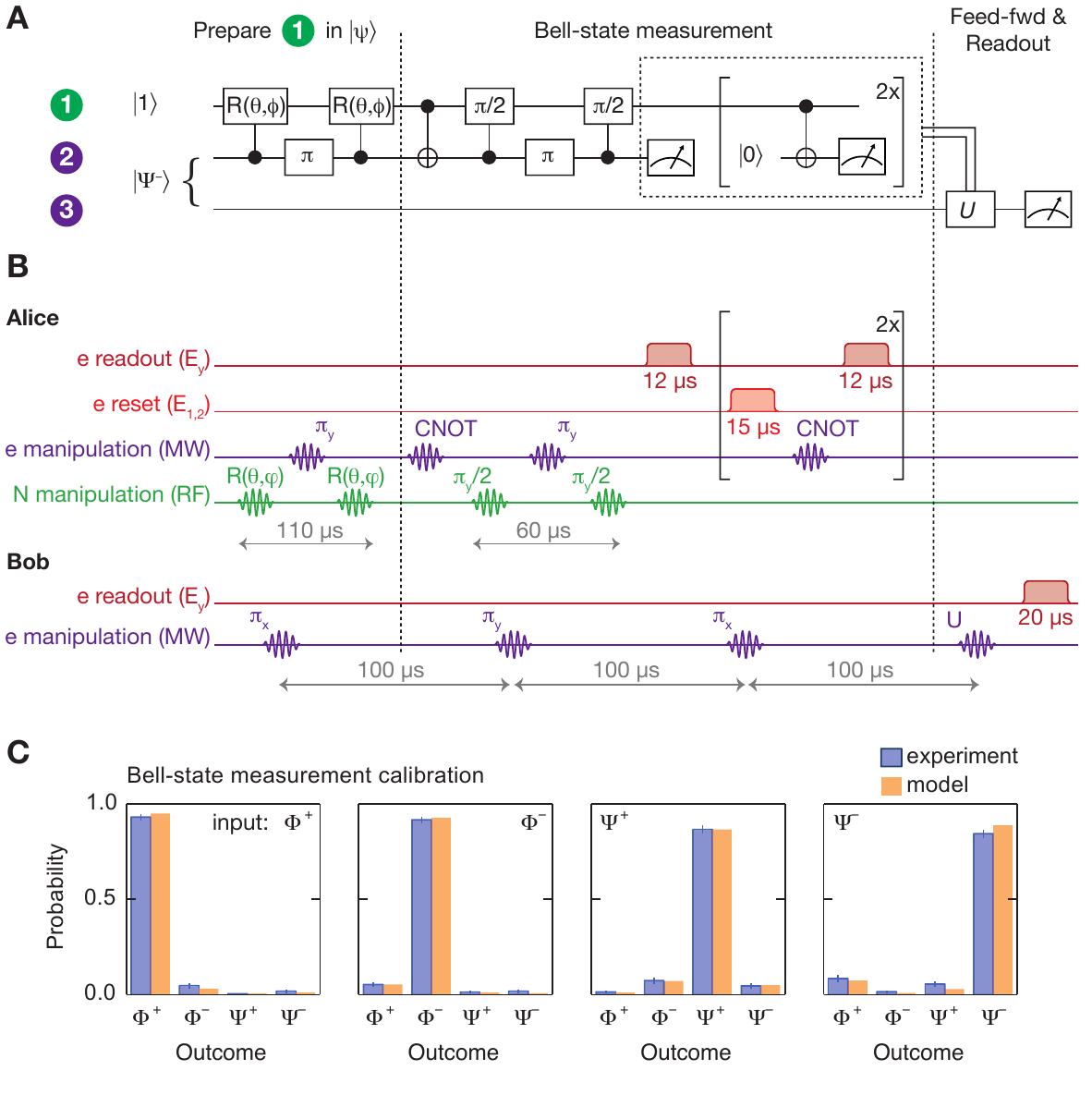}
    \caption{
    \label{fig3} 
    Deterministic Bell-state measurement (BSM) and real-time feed-forward. 
    \textbf{(A)} Circuit diagram and \textbf{(B)} pulse scheme of our implementation. The label `e' (`N') indicates operations acting on the electron spin (nitrogen nuclear spin). To enhance the readout fidelity for the nuclear spin, we perform the mapping to the electron spin via a CNOT and the subsequent electron readout twice. While Alice is performing her BSM Bob applies an XY4 decoupling sequence on his electron qubit. After receiving the BSM outcome from Alice, Bob applies the feed-forward operation $U$ and reads out his qubit. $\pi_{x,y}$ denote rotations around the $x$-axis and $y$-axis, respectively. 
    \textbf{(C)} Calibration of the BSM by inserting the four different Bell states on Alice's side and determining the probability with which the ideal outcome is observed (blue bars). Data is not corrected for imperfect preparation of the input states. Expectations based on independently determined experimental imperfections are shown in orange. Error bars are two statistical s.d.
    }
\end{figure*}

The additional qubit in Alice's node --- essential for making the teleportation unconditional --- is provided by the nitrogen-14 nuclear spin of Alice's NV (qubit 1). Before establishing the entanglement link, this nuclear spin is initialized into state $\ket1$ by a projective measurement via the electron spin~\cite{Robledo:2011fs}. We reinitialize the nuclear spin after each 250 entanglement attempts in order to preserve its purity (Figs.~\ref{fig2}C,D). We prepare the source state after establishing remote entanglement, thus avoiding possible dephasing of the source state by repeated optical excitation of the nearby electron~\cite{2008PhRvL.100g3001J,2014NatPh..10..189B} during entanglement generation. We employ a decoherence-protected gate~\cite{vanderSar:2012if} on Alice's side to set the nuclear spin to the source state $\ket\psi = \alpha\ket0 + \beta\ket1$. This gate combines two nuclear spin rotations with a refocusing pulse on the electron spin such that the entangled state is efficiently preserved for the duration of the gate (Figs.~\ref{fig3}A,B). This operation concludes the preparation of the teleporter and the insertion of the source qubit, with the three-qubit system left in the state $\ket\psi_1 \otimes \ket{\Psi^-}_{23} = (\alpha\ket0_1 + \beta\ket1_1) \otimes (\ket{01}_{23} - \ket{10}_{23})/\sqrt2$.

At the heart of unconditional qubit teleportation is a deterministic Bell-state measurement (BSM) by Alice on qubits 1 and 2 that generally involves two steps. First, the four Bell states are mapped onto the four different qubit eigenstates $\ket{i}_1 \ket{j}_2$ by quantum gate operations. In the second step each of the two qubits is read out in a single shot and the two measurement outcomes are sent to Bob. Our implementation of this scheme is shown in Figs.~\ref{fig3}A and B. We implement the Bell-state mapping by applying a two-qubit controlled-NOT gate (CNOT) followed by a $\pi/2$ rotation on the nuclear spin using another decoherence-protected gate. Then we read out the electron spin in a single shot (average fidelity $0.963\pm0.005$). Finally we read out the nuclear spin by mapping its state onto the electron spin followed by electron spin readout. The two single-shot readout results give the outcome of the BSM.

We benchmark the BSM by preparing each of the four Bell states as input states in Alice's register (Fig.~\ref{fig3}C). This procedure yields an uncorrected mean fidelity, given by the probability to obtain the measurement result corresponding to the prepared Bell state, of $0.89\pm0.02$. To gain more insight into the sources of imperfections we compare the data with numerical simulations that use the independently determined infidelities of the nuclear spin initialization, CNOT gate, and electron single-shot readout as input. These simulations predict an average fidelity of 0.9 (Fig.~\ref{fig3}C), in excellent agreement with the data. Taking known errors in the preparation of the input states into account, we infer a BSM fidelity of $0.93\pm0.02$.

The final challenge for successful unconditional teleportation is to maintain the coherence of Bob's target qubit (qubit 3) during the BSM and feed-forward. In our experiment, Bob's qubit is mostly affected by interactions with the surrounding nuclear spin bath. We counteract this decoherence by applying an XY4 dynamical decoupling sequence~\cite{2010Sci...330...60D}. The time between entanglement generation and the triggering of the feed-forward operation based on the BSM outcome is 300\,\textmicro s. For this duration the decoupling protocol preserves the qubit state with an average fidelity of $0.96\pm0.02$.

\begin{figure*}
    \includegraphics{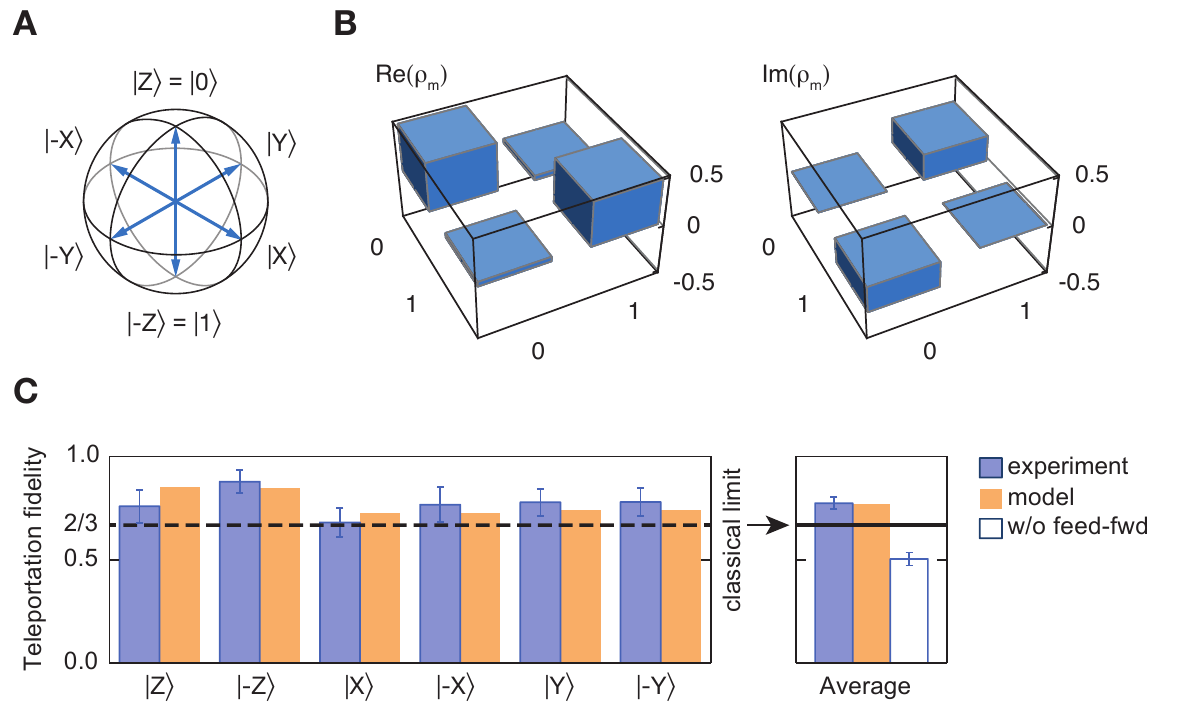}
    \caption {
    \label{fig4} 
    Demonstration of unconditional quantum teleportation between remote qubits. 
    \textbf{(A)} Bloch sphere with the six mutually unbiased basis states that we teleport. $\ket{\pm X} = (\ket0 \pm \ket1)/\sqrt2$, $\ket{\pm Y} = (\ket0 \pm \ii \ket1)/\sqrt2$.
    \textbf{(B)} State tomography after teleportation of the input state $\ket Y$. We determine the density matrix $\rho_\mathrm m$ by measuring the expectation values of the Pauli spin operators, $\langle \sigma_x \rangle$, $\langle \sigma_y \rangle$, $\langle \sigma_z \rangle$, where the required qubit rotations before readout are performed conditional on the BSM outcome. The measured (ideal) entries of the density matrix are $\rho_{00} = 1 - \rho_{11} = 0.52 \pm 0.08 \; (0.5)$ and $\rho_{01} = \rho_{10}^* = 0.05 \pm 0.08 - \ii 0.28 \pm \ii 0.07\; (- \ii 0.5)$, respectively.
    \textbf{(C)} Average teleportation fidelity from the measured fidelities of the six states (blue bars). Sample sizes are (left to right) 54, 89, 73, 49, 52, and 47. Predictions from simulations are shown in orange. Without feed-forward, the target state is completely mixed (white bar). The horizontal line marks the classical limit of $2/3$. Data is not corrected for source state initialization errors. Uncertainties are one statistical s.d. 
    }
\end{figure*}

We first verify that the teleporter is calibrated correctly by applying it to the nominal input state $\ket Y = (\ket0+\ii\ket1)/\sqrt2$ and performing tomography on the state that appears on Bob's side. The reconstructed density matrix (Fig.~\ref{fig4}B) shows that the target state vector is aligned well with $Y$ and therefore that the reference frames of Alice and Bob are correctly set.

To prove that our quantum teleporter outperforms any classical communication strategy, we teleport an unbiased set of six basis states $\ket\psi$ (Fig.~\ref{fig4}A) and determine the fidelity of the teleported state on Bob's side with respect to the ideal input state. In these experiments we use a feed-forward operation that maps the ideal state of qubit 3 onto a qubit eigenstate such that the readout directly yields the teleportation fidelity. Since the feed-forward operation is conditional on the BSM outcome, ignoring the BSM outcome yields a completely mixed state and random outcomes ensuring that no information is transmitted. Without feed-forward we indeed observe an average teleportation fidelity of $\langle F \rangle = 0.50 \pm 0.03$ (Fig.~\ref{fig4}C). In contrast, including the feed-forward loop we find $\langle F \rangle = 0.77 \pm 0.03$. This value exceeds the classical limit of $2/3$ by more than 3 standard deviations, thus proving the quantum nature of our teleporter. We note that this fidelity presents a lower bound on the actual teleportation fidelity because it does not take into account initialization errors of the source state. Importantly, this result is obtained without any post-selection: each teleportation attempt is included in the data presented here.

We also simulate the outcomes by using independently determined infidelities in the protocol. The only unknown parameter is the fidelity of the entangled state shared by Alice and Bob. We find that our data is well reproduced by the simulations if we assume a fidelity to the ideal Bell state $\ket{\Psi^-}_{23}$ of 0.87 (Fig.~\ref{fig4}C). The simulations also enable us to quantify the effect of imperfect initialization of the source qubit on the measured fidelities. In this way we estimate the teleportation fidelity to be $\sim 0.86$.

The ability to generate remote entanglement and to control and read out multiple qubits per node as shown in the present teleportation experiment makes NV centers a leading candidate for realizing a quantum network. Our teleportation scheme is both unconditional and scalable to large distances as it can mitigate photon loss by heralding and purification of the distributed entangled state~\cite{1998PhRvL..81.5932B}. In future experiments we aim to supplement our current capabilities with quantum memories that are robust against optical excitation of the electrons, enabling remote entanglement purification and the connection of multiple nodes into the network. A promising route is the use of weakly coupled nuclear spins on which multi-qubit quantum control has very recently been demonstrated~\cite{2013arXiv1309.5452T}. For such nuclear spins, coherence times of over 1 second under optical excitation have been reported~\cite{Maurer:2012kg}, while the incorporation of NV centers into optical cavities may enable remote entanglement generation on millisecond timescales~\cite{Loncar:2013kv}. Furthermore, the entanglement and readout fidelities reported here are sufficient for a violation of a Bell inequality with the detection loophole closed, making NV centers a promising system for realizing a loophole-free Bell test and device-independent quantum key distribution~\cite{2013arXiv1303.2849B}. 

\graphicspath{{./SOM_figures/}}

\setcounter{figure}{0}
\makeatletter 
\renewcommand{\thefigure}{S\@arabic\c@figure} 

\setcounter{table}{0}
\makeatletter 
\renewcommand{\thetable}{S\@Roman\c@table} 

\setcounter{equation}{0}
\makeatletter 
\renewcommand{\theequation}{S\@arabic\c@equation} 

\section{Materials and methods}

\subsection{Experimental techniques}

\subsubsection{Samples}

\begin{figure*}
    \includegraphics{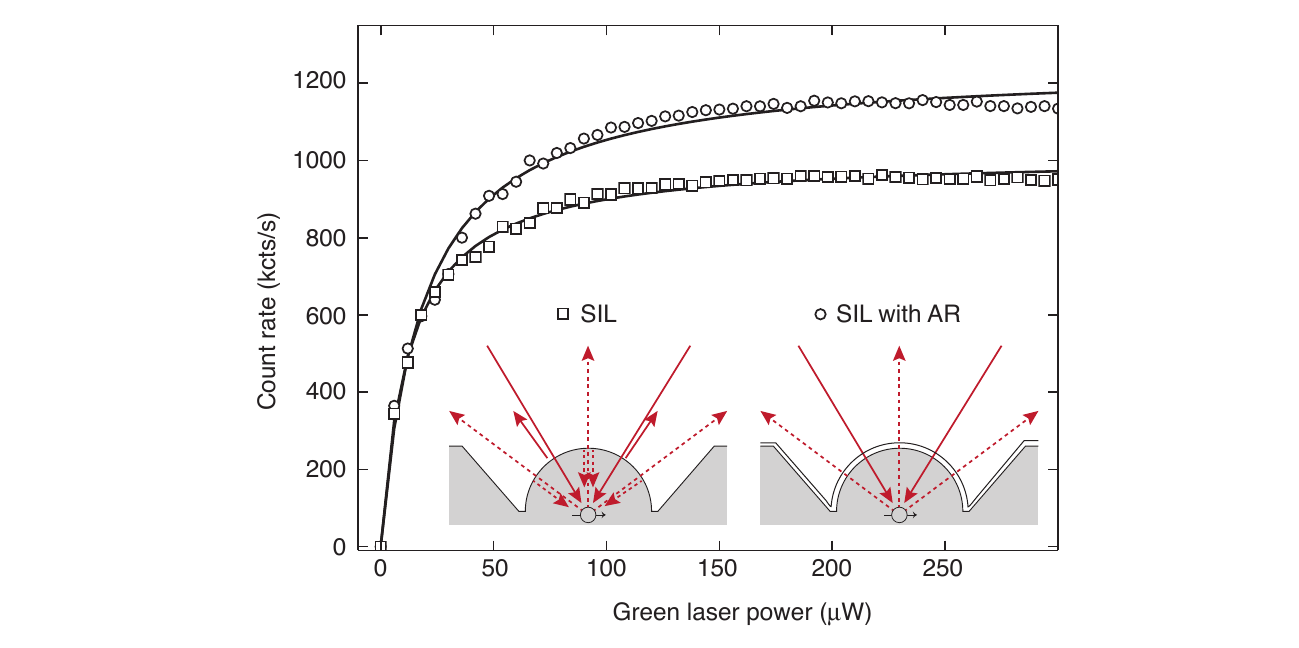}
    \caption{
    \label{fig:AR-SIL}
    Saturation measurements on SILs with and without antireflection coating. 
    Fluorescence count rates in dependence of off-resonant green excitation power (kcts = 1000 counts). Solid lines are fits to $A \cdot x / (x + P_\text{sat})$. In the case of a bare SIL, photons emitted from the NV centre and the excitation laser can be reflected at the interface due to the large refractive index of diamond. This effect is overcome by an antireflection coating which further increases the count rates and significantly reduces reflections of the excitation laser.
    }
\end{figure*}

We use naturally occurring Nitrogen-Vacancy (NV) centres in high purity type IIa chemical-vapor deposition grown diamond with a $\langle111\rangle$ crystal orientation obtained by cleaving a $\langle100\rangle$ substrate. We select NV centres that are aligned along the $\langle111\rangle$ direction and that do not couple strongly to any \cthirteen-nuclear spins in the vicinity. Around these preselected centres we deterministically fabricate solid immersion lenses (SILs) in order to enhance the collection efficiency \cite{2010ApPhL..97x1901H,2011ApPhL..98m3107M,Robledo:2011fs}. On the surface of Alice's diamond we have additionally deposited a single-layer antireflection coating made from aluminum oxide that has been designed for best performance at a wavelength of 637\,nm \cite{2012ApPhL.100y1111Y}. This coating increases the collection efficiency (see Figure~\ref{fig:AR-SIL}). Furthermore, it significantly reduces reflections from the resonant excitation laser.

\subsubsection{Setup}

The experimental setups used are similar to the one used in Bernien {\em et al}.~\cite{2013Natur.497...86B}. We perform the experiments with two home-built low-temperature confocal microscopes. Alice's sample is mounted on a XYZ stepper/scanner piezo stack (Attocube) in a Janis ST-500 flow cryostat and kept at $T \approx 8$\,K. Bob's sample is mounted on a XYZ stepper (Attocube) inside a custom-built Cryovac bath cryostat with optical access and kept at a temperature of $T \approx 4$\,K. Each setup features lasers for off-resonant and resonant excitation, cryogenic piezoelectric positioners and high-efficiency/low background fluorescence detection paths. The zero-phonon line (ZPL) detection paths of both setups lead to a common beam splitter and photon detectors used for the remote entanglement generation. 

Off resonant green excitation is provided for each of the setups by 532\,nm lasers (Spectra Physics Millenia Pro and Laser 2000 Cobalt Samba for Alice and Bob, respectively). Alice additionally features yellow excitation at 575\,nm from a frequency-doubled diode-laser (Toptica). Two tuneable 637\,nm lasers (New Focus Velocity) for independent resonant excitation are used for optical spin-pumping. All lasers are pulsed by acousto-optic modulators (AOMs; Crystal Technologies). 

For the resonant excitation pulses used to generate the entanglement both setups share a tuneable continuous wave 637\,nm laser (Sirah Matisse DS). Its output is sequentially fed through an AOM and an electro-optic modulator (EOM; Jenoptik). After passing through the AOM \& EOM, the beam is split using a 50/50 beam splitter, and a 30 cm adjustable delay line is inserted in one arm for fine-tuning the temporal overlap of the excitation.

The photon emission of each NV is split into a ZPL part and an off-resonant phonon sideband (PSB) part by a dichroic long-pass filter (Semrock LPD01-633RS). The PSB emission is independently detected for each setup by avalanche photo-diodes (APDs; Perkin-Elmer SPCM). The ZPL emission is further filtered by a second dichroic filter (to remove green excitation light) and a tuneable band pass filter (Semrock TBP-700B). After filtering resonant excitation light by cross-polarisation rejection the ZPL emission of NVs A and B is coupled into the input ports of a fibre-coupled beam splitter (Evanescent Optics) by polarisation-maintaining fibres. The photons leaving the output ports of the beam splitter are detected by fibre-coupled avalanche photo-diodes (Picoquant Tau-SPAD) and time-tagged by a Picoquant Hydraharp 400 system. 

All laser frequencies are monitored by a high-precision wave meter (High Finesse Angstrom WSU) and stabilized by software feedback using DAC modules.

The control signals that generate the optical pulses on the AOMs and EOM synthesized by two Arbitrary Waveform Generators (AWGs; Tektronix AWG 5014C), each operating on one setup. The joint path for entanglement generation is controlled by Bob's AWG. The AOMs also can be controlled simultaneously via DAC modules (see below).

Each setup has an independent microwave (MW) source (Rohde \& Schwarz SMB100A) and MW amplifier (Amplifier Research 20S1G4 and AR 40S1G4 for the setups of Alice and Bob, respectively) to drive the NV centre electron spins. Alice's nuclear spin is driven by RF control fields synthesized directly in the AWG and fed through an RF amplifier (Electronic \& Innovation 240L) before combining with the MW line.

Gate voltages for tuning of optical resonances by the d.c.\ Stark effect are controlled via a DAC module and amplified by a home-built DC amplifier.

\subsubsection{Protocol implementation}

\begin{figure*}[tp]
    \includegraphics{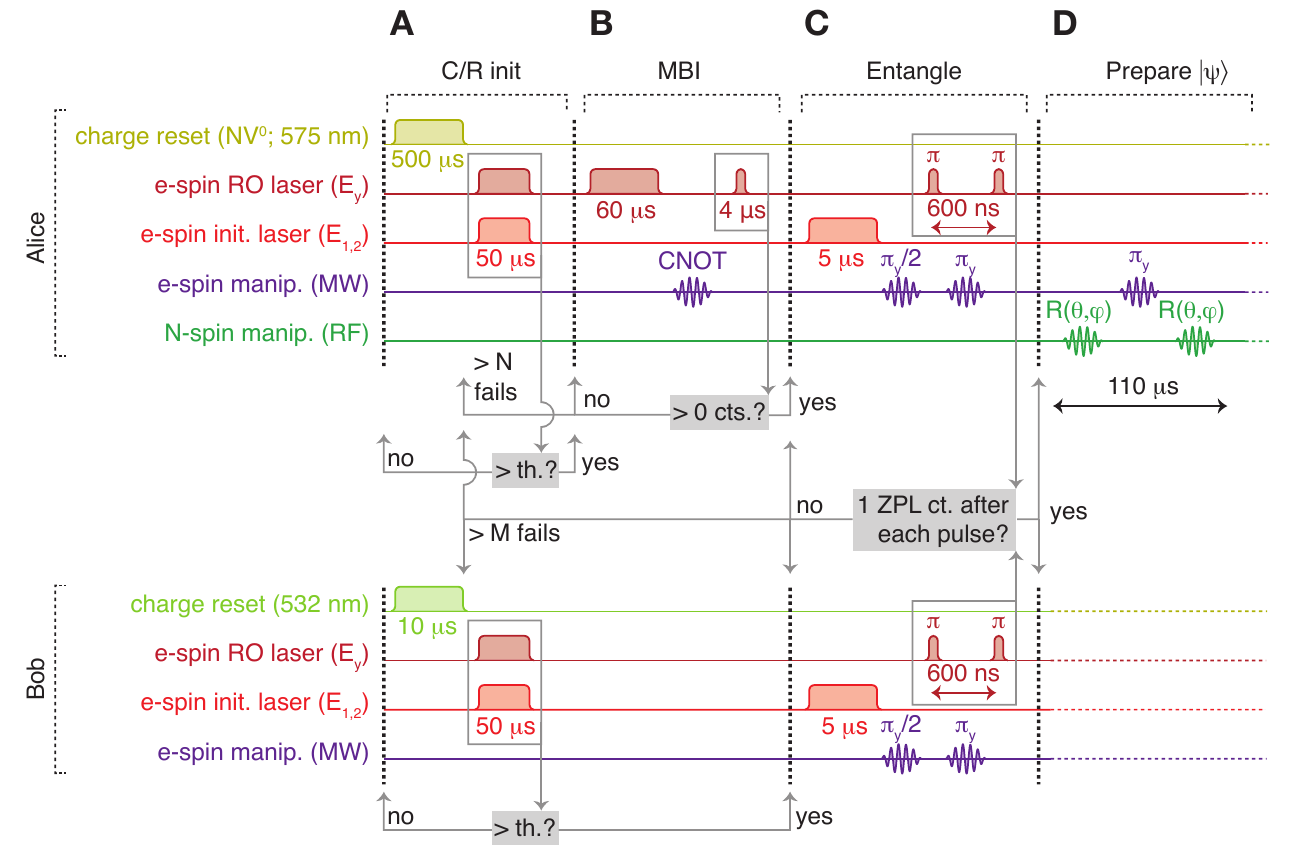}
    \caption{
    \label{fig:teleportation-init} 
    System initialization.
    \textbf{(A)} We verify charge and resonance condition of Alice and Bob (asynchronously) by applying laser pulses on $E_y$ and $E_{1,2}$ simultaneously and putting a lower threshold on the number of phonon side band photons detected during those pulses. If the threshold is not met we reset the charge state: On Alice, we repump NV$^0$ $\rightarrow$ NV$^-$ using a laser at 575\,nm, on resonance with the ZPL of NV$^0$~\cite{Siyushev:2013en}. On Bob, we use off-resonant excitation at 532\,nm. We repeat verification and repump until success.
    \textbf{(B)} Following spin-pumping into $\mspmone$ by excitation of $E_y$ we apply a CNOT on the electronic spin, such that rotation to $\mszero$ is only performed for $\mimone$. A PSB photon detected during a short readout pulse on $E_y$ signals a high-fidelity measurement of $\mszero$ and projection of the nuclear spin into $\mimone$. If no photon is detected, we re-try for a maximum of $N$ times (here, $N=100$), before charge and resonance are re-evalutated. In between attempts we apply $50\,\mu\mathrm s$ of illumination on both $E_y$ and $E_{1,2}$ in order to randomise the nuclear spin owed to off-diagonal terms in the hyperfine interaction in the optical excited state (not shown in the diagram).
    \textbf{(C)} As soon as both Alice and Bob are initialised, we attempt to generate entanglement between them. Each attempt starts with an electron spin reset to $\mszero$. Two rounds of optical excitation with optical $\pi$-pulses on $E_y$ follow, separated by a MW $\pi$-pulse. Detection of exactly one ZPL photon after each pulse heralds creation of entanglement. We perform a maximum of $M$ attempts before re-initialisation (here, $M=250$). 
    \textbf{(D)} When entanglement is created, we prepare the \nfourteen\ spin of Alice unconditional on the electron spin state, while preserving the electron spin phase. The RF pulse that generates the rotation is only resonant for $\msmone$; we perform the rotation twice, separated by a $\pi$-pulse on the electron.
    }
\end{figure*}

In order to maintain a high repetition rate of the experiment we implement a conditional protocol as follows. We first ensure that both NVs are in their negatively charged state and that our lasers are on resonance with the optical transitions used (Fig.~\ref{fig:teleportation-init}A)~\cite{Robledo:2011fs,2013Natur.497...86B}. 

The next step is to create entanglement between the the electronic spins and to prepare Alice's nuclear spin in the source state. Each attempt to generate entanglement between the electronic spins consists of preparation into $\ket 0$ by optical spin-pumping followed by two rounds of spin-rotation by microwaves and spin-selective optical excitation (Fig.~\ref{fig:teleportation-init}C), conducted simultaneously on Alice's and Bob's sides~\cite{2005PhRvA..71f0310B,2013Natur.497...86B}. Each attempt takes on the order of 10\,\textmicro s and results in heralded entanglement with a success probability of $\sim 10^{-7}$. 

To avoid disturbance of Alice's nuclear spin we keep it in an eigenstate, created by measurement-based initalisation, during entanglement generation (Fig.~\ref{fig:teleportation-init}B)~\cite{Robledo:2011fs,2013NatPh...9...29P}. We preserve the purity of the state by re-initialising after every 250 failed entanglement generation attempts (see below).

As soon as the entanglement generation succeeds, we prepare the source state $\ket\psi$ (Fig.~\ref{fig:teleportation-init}D) using a decoherence protected gate~\cite{vanderSar:2012if}.

The conditional operation described is implemented in the following manner: The charge and resonance (CR) check and readout is done independently for the two setups by two programmable micro-controllers with DAC- and counter modules (Adwin Gold II and Adwin Pro II for setup A and B, respectively) that can control the AOMs. Alice's Adwin further prepares the nuclear spin by measurement by first performing spin-pumping using the $E_{1,2}$ AOM, triggering the CNOT sequence on the AWG, and finally reading out the electron spin optically using the $E_y$ AOM.

Once this initialization is finished, a start trigger is sent to the AWGs that sequentially execute the entanglement protocol. For each round, the photon clicks are recorded and time-tagged by a Picoquant Hydraharp 400 system. In addition, the photon clicks are also monitored in real time by a programmable logic device (CPLD; Altera Max V development kit) that time-filters the signal and recognises a successful entanglement event; if a success occurs, a trigger is sent to the AWGs, starting the BSM and dynamical decoupling sequences. 

The two electron spin readouts of the BSM are performed by Alice's Adwin. Conditional on the readout result, a feed-forward sequence on Bob is triggered, before Bob's Adwin performs single-shot readout of the target state.

\section{Supplementary Text}

\subsection{Teleportation protocol}

\subsubsection{Conventions}

The basis states used for the electrons are $\ket{0} = \ket{\mszero}$ and $\ket{1} = \ket{\msmone}$. For the nitrogen, $\ket{0} = \ket{\mizero}$ and $\ket{1} = \ket{\mimone}$. When specifying joint quantum states, the first qubit is the nitrogen on site A, the second the electron on site A, and the third the electron on site B. Teleportation is performed from qubit 1 onto qubit 3.

By $x, y, z$ we denote $\pi/2$ rotations around the $+X, +Y, +Z$ axes respectively. Bars over gate symbols indicate negative rotation sense. In the measurement sequences, rotations around $+X, +Y$ correspond to phases of the applied driving pulses of $+90^\circ$ and $0^\circ$, respectively. We prepare $\ket{x} \equiv (\ket{0}+\ket{1})/\sqrt2$ by $y \ket{0}$ and $\ket{y} \equiv (\ket{0}+\mathrm i \ket{1})/\sqrt2 = \bar{x}\ket{0}$. Capital letters $X, Y, Z$ indicate $\pi$ rotations.

\subsubsection{Hamiltonian of Alice}

\begin{figure*}[htpb]
    \centering
    \includegraphics{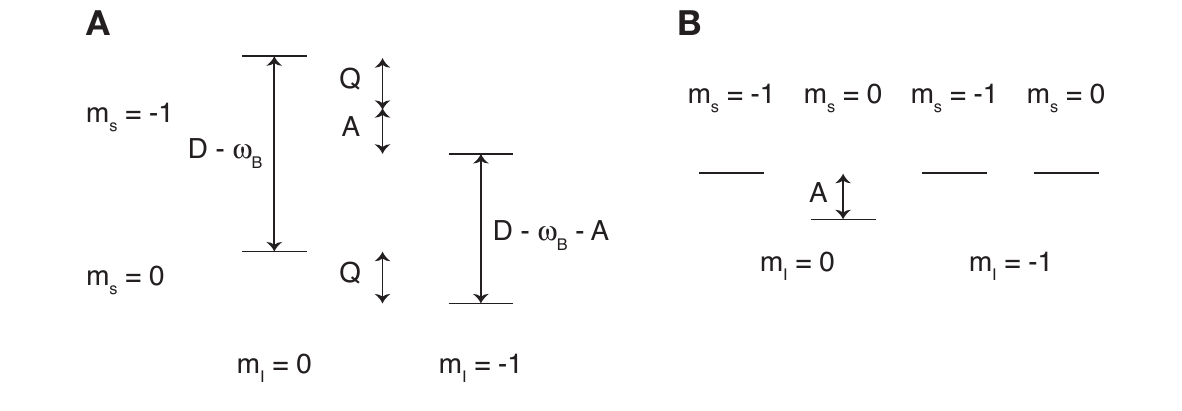}
    \caption{
    \label{fig:teleportation-SOM_system-Alice}
    Relevant spin states on Alice's side.
    \textbf{(A)} Lab frame. 
    \textbf{(B)} Rotating frame chosen. $D = 2\pi \times 2.878\,\text{GHz}$ is the NV electron zero-field splitting, $\omega_\mathrm B \approx 2\pi \times 50$ MHz is the Zeeman splitting of the electron, $A = 2\pi \times 2.19$ MHz is the electron-nitrogen hyperfine coupling constant.}
\end{figure*}

The relevant energy levels of the electron and nuclear spins of Alice are depicted in Fig.~\ref{fig:teleportation-SOM_system-Alice}a. We chose the rotating frame (Fig.~\ref{fig:teleportation-SOM_system-Alice}b) such that the relevant Hamiltonian without driving can be written as
\be
    \mathcal{H}^\mathrm A_0 = 
    \begin{pmatrix}
        -A & 0 & 0 & 0 \\
        0 & 0 & 0 & 0 \\
        0 & 0 & 0 & 0 \\
        0 & 0 & 0 & 0
    \end{pmatrix},
\ee
where $A = 2\pi \times 2.19$\,MHz is the parallel hyperfine coupling constant of electron and nitrogen at low temperatures. The spin eigenstates are $\ket{00}, \ket{01}, \ket{10}, \ket{01}$.

\subsubsection{Desired state evolution}

\paragraph{Source state preparation}

After generating entanglement, we start with the state $\ket{1} \left ( \ket{01} - \ket{10} \right ) / \sqrt{2}$.
We perform the desired rotation on the nitrogen spin for the $\msmone$ manifold, then apply a $\pi$-pulse to the electron and repeat the operation. In this way the operation on the nitrogen spin is unconditional on the electron state and the electron phase is protected by a spin-echo. With an RF operation $\ket1 \mapsto \alpha\ket0 + \beta\ket1$ this procedure yields

\be
    \frac{1}{\sqrt 2} \left( 
        \left(\expe^{-\ii A(t-t_0)} \alpha \ket0 + 
            \beta \ket1 \right) \ket{00}
        + \left( \alpha \ket0 + \beta \ket1 \right) \ket{11}
    \right).
\ee

Note that the states associated with $\ket{00}$ on Alice's side accumulate a phase during free evolution time, $t$, due to the choice of rotating frame. $t_0$ is the time at which the $\pi$-pulse on the electron is performed during preparation. By chosing the evolution time such that $A(t-t_0)$ is a multiple of $2\pi$ the initial state can be factorized.

We implement the unconditional rotation of the electron spin with a CORPSE pulse that provides a $\pi$ rotation that is insensitive against detuning over a a range of a few MHz~\cite{2003PhRvA..67d2308C}.

\paragraph{Bell-state measurement}

The BSM consists of a CNOT rotation around the $+Y$ axis on Alice's electron spin, conditional on the nitrogen spin being in $\ket0$, followed by a $\pi/2$ rotation around the $+Y$ axis on the nitrogen spin. We implement the CNOT by rotating $\mimone$ by $\pi$ and $\mizero$ by $2\pi$, achieved by a pulse with Rabi frequency $A/\sqrt{3}$. During this pulse Alice's states $\ket{00}$ and $\ket{01}$ are not unaffected. In particular, the time-dependent phase of the state $\ket{00}$ is reduced compared to not performing the pulse (or compared to the case of an ideal CNOT gate in which only a real $\mathbb{1}$ operation would be applied to this state) because some population temporarily leaves this state. Conversely, $\ket{01}$ will acquire some phase because some population will temporarily be in $\ket{00}$. An unconditonal rotation of the nitrogen spin is achieved in the same was as for preparation, by performing the operation twice, with an electron flip in between. After these gate operations we have
\begin{align}
    \frac{1}{2} \bigg[
          & \ket{00} \left( \beta\ket0 - 
            \expe^{\ii\lambda}\alpha\ket1 \right) \notag\\
        + & \ket{01} \left( \expe^{-\ii A(t_1-t_0)-\ii\kappa}\alpha\ket0 +
            \beta\ket1 \right) \notag\\
        + & \ket{10} \left( -\beta\ket0 - 
            \expe^{\ii\lambda}\alpha\ket1 \right) \notag\\
        + & \ket{11} \left( \expe^{-\ii A(t_1-t_0)-\ii\kappa}\alpha\ket0 -
            \beta\ket1 \right)
    \bigg],
\end{align}
where $t_1$ is the time of the $\pi$-pulse on the electron and $\lambda,\kappa$ are the additional phases on $\ket{00}$ and $\ket{01}$.

\paragraph{Phase calibration}

\begin{figure*}[htp]
    \centering
    \includegraphics{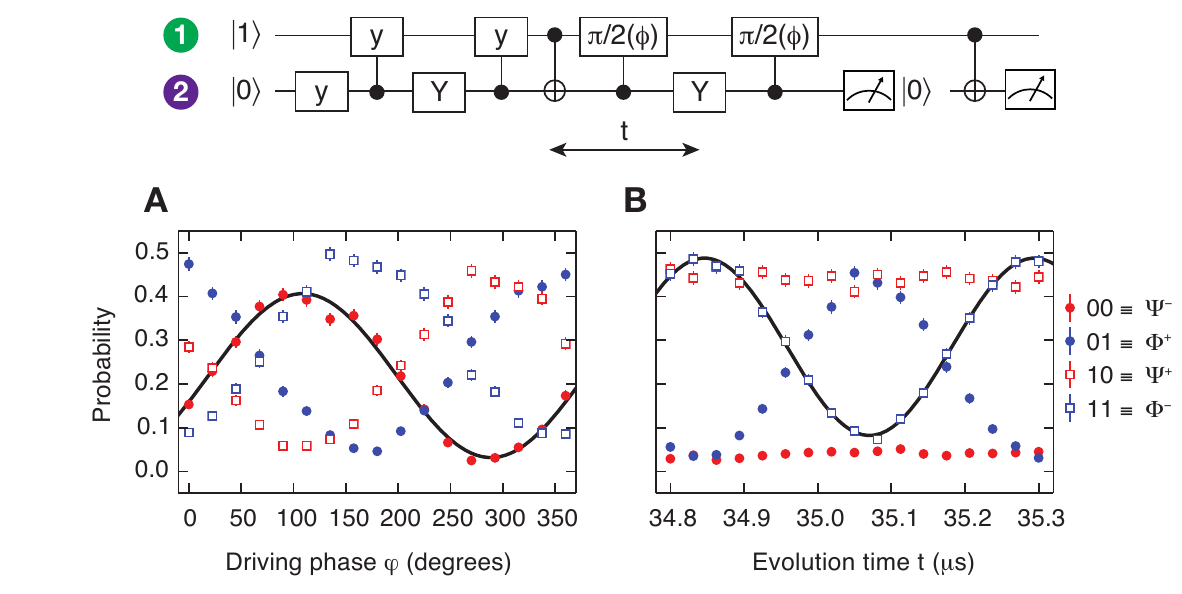}
    \caption{
    \label{fig:teleportation-SOM_BSM-calibration}
    Calibration of the Bell-state measurement.
    \textbf{(A)} Calibration of the driving phase of the Hadamard operation, and 
    \textbf{(B)} subsequent calibration of the evolution time between the CNOT gate of the BSM and the electron $\pi$-pulse for the unconditional rotation of the nuclear spin. The solid lines are sinosoidal fits to the BSM outcomes to be maximised. The legend indicates the correspondence between two-qubit measurement results $ij$ and Bell-state detection. The calibration is performed with the full teleportation protocol including the MW pulses during entanglement generation attempts (but without optical $\pi$-pulses). Error bars are 1 s.d.}
\end{figure*}

We can eliminate the undesired phases before the teleportation experiment by calibrating the rotation axis of the $\pi/2$ operation on the nitrogen in the BSM and the evolution times. After initializing the nitrogen and electron spin states of Alice into $\ket{1}(\ket{0} - \ket{1})/\sqrt{2}$ (equivalent to the entanglement operation on Alice, ignoring Bob), we prepare the nitrogen in $\ket{\bar x} = (\ket{0} - \ket{1})/\sqrt{2}$ (preparation operation is $y$) and perform the BSM, yielding
\begin{align}
    \frac{1}{2\sqrt2} \bigg[
          & \ket{00} \left( -1 - \expe^{\ii\lambda} \right) \notag\\
        + & \ket{01} \left( -1 + 
            \expe^{-\ii A(t_1-t_0) -\ii\kappa} \right) \notag\\
        + & \ket{10} \left( 1 - \expe^{\ii\lambda} \right) \notag\\
        + & \ket{11} \left( 1 + \expe^{-\ii A(t_1-t_0) -\ii\kappa} \right)
    \bigg]
\end{align}
before readout (Fig.~\ref{fig:teleportation-SOM_BSM-calibration}). We sweep the rotation axis of the RF pulse on the nitrogen (affecting the phase $\ii\lambda$) and subsequently the evolution time between the CNOT and $Y$ operations during the BSM (affecting the phase $-\ii A(t_1-t_0) -\ii\kappa$). Calibration is achieved by maximizing the probabilities for outcomes $\ket{00}$ and $\ket{11}$. 

\paragraph{Dynamical decoupling of Bob's electron spin}

To protect the target state against dephasing during the BSM, we perform an XY4 decoupling sequence in parallel. The first $\pi$-pulse of this echo sequence is the $\pi$-pulse performed during the entanglement generation attempt. The remaining X-Y-X sequence is executed during the BSM. Taking these additional rotations into account, the total state before readout, including phase calibration, is
\begin{align}
    \label{eq:BSM-outcomes}
    \frac{1}{2} \bigg[
          & \ket{00} \left( \alpha\ket0 + \beta\ket1 \right) \notag\\
        + & \ket{01} \left( -\beta\ket0 + \alpha\ket1 \right) \notag\\
        + & \ket{10} \left( \alpha\ket0 - \beta\ket1 \right) \notag\\
        + & \ket{11} \left( \beta\ket0 + \alpha\ket1 \right)
    \bigg].
\end{align}

Because we do not intialize the nuclear spin on Bob's side we perform all electron spin rotations with CORPSE pulses~\cite{2003PhRvA..67d2308C}.

\paragraph{Feed-forward}
The required feed-forward operations to re-create $\ket{\psi}$ on the target spin can be taken straight-forward from Eq.~\ref{eq:BSM-outcomes}. For the estimation of the fidelity of the teleported state with the source state it is sufficient to read out in the basis aligned with the source state vector. We achieve this readout by modifying the feed-forward operation such that we rotate the target state $U_{i,j}\ket \psi$ directly into the $z$-basis, conditional on the outcome of the BSM. The operations we apply in practice are summarized in Table \ref{tab:psiminus-operations}.

\begin{table}[htp]
    \centering
    \caption{
    \label{tab:psiminus-operations}
    Feed-forward and readout operations applied for each BSM outcome.}
    \vspace{.2cm}
    \begin{tabular}{l | c c c c || l}
        Input & $\ket{00}$ & $\ket{01}$ & $\ket{10}$ & $\ket{11}$ & ideal result \\
        \hline
        $\ket{+z} = Y\ket{1}$ & $\mathbb 1$ & $Y$ & $\mathbb 1$ & $Y$ & $\ket{0}$ \\
        $\ket{-z} = \mathbb 1 \ket{1}$ & $Y$ & $\mathbb 1$ & $Y$ & $\mathbb 1$ & $\ket{0}$ \\
        $\ket{+x} = \bar y \ket{1}$ & $\bar y$ & $y$ & $y$ & $\bar y$ & $\ket{0}$ \\
        $\ket{-x} = y \ket{1}$ & $y$ & $\bar y$ & $\bar y$ & $y$ & $\ket{0}$ \\
        $\ket{+y} = \bar x \ket{1}$ & $\bar x$ & $\bar x$ & $x$ & $x$ & $\ket{1}$\\
        $\ket{-y} = x \ket{1}$ & $x$ & $x$ & $\bar x$ & $\bar x$ & $\ket{1}$\\
    \end{tabular}
\end{table}

\subsection{Data analysis}

For each input state we determine the number of events $n_{0}$ and $n_{1}$ that give measurement outcomes $\mszero$ and $\msmone$, respectively. The probability amplitudes $c_0$ and $c_{1}$ for $\ket0$ and $\ket1$ are obtained by performing readout correction using the readout fidelities $F_0$ and $F_{-1}$ for $\mszero$ and $\msmone$, respectively. We obtain $F_0$ and $F_{-1}$ from calibration measurements performed periodically during the experiment. The teleportation fidelity of the state is given by either $c_0$ or $c_{1}$ (see Table~\ref{tab:psiminus-operations}).

The uncertainty of $c_0$ and $c_1$ is determined by the standard deviation of the binomial distribution with probabilities $n_0/(n_0 + n_1)$ and $n_1/(n_0 + n_1) = 1 - n_0/(n_0 + n_1)$, and the measurement uncertainties of $F_0$ and $F_{-1}$ (for both readout fidelities the measurement uncertainties are $\lesssim 0.005$).

\subsection{Error model}

In the following we describe the errors we take into account for modeling our experimental results. Any further errors are considered small in comparison and we ignore them in this discussion. In particular we assume that the preparation of the source state $\ket\psi$ is not subject to errors resulting from RF or microwave pulses.

Note that we model the experimental results numerically with the best guesses of the empiric parameters described below, without treatment of their uncertainties.

In general we simulate the experimental results by modeling the system by a $2\times 2 \times 2$ dimensional density matrix that is subjected to operators that correspond to the operations physically applied. Treatment of errors is included in the description of the types of errors taken into consideration in the following. Operations for which no error is listed are assumed to be perfect.

\subsubsection{CNOT pulses}

The fidelity of Alice's electron spin rotations that are selective on the \nfourteen\ spin state are limited by the finite linewidth of the electron spin transitions. We simulate the effect of the pulse on the different hyperfine populations by evaluating the probability for inversion versus detuning using a master equation solver \cite{2013CoPhC.184.1234J} and integrating over the transition line shapes. In this way we compute the probabilities for an erroneous inversion for $m_\mathrm I = -1$ and non-inversion for $m_\mathrm I = 0$ to be both 0.01. Our calculation is based on a finite linewidth that is determined by the electron spin dephasing time, $T_2^* = 2\,\mu\mathrm s$. In our model we assume that in case of an error the spin state is dephased (i.e., we numerically set the respective coherences in the resulting density matrix to zero).

\subsubsection{Nuclear spin initialization}

When preparing the source state to be teleported, the following errors can occur: (1) Initialization by measurement into $\mimone$ succeeds with a fidelity $p_{-1}$, and fails for the initial state in either $\mizero$ or $\mipone$, with probabilities $p_0$ and $p_{+1}$, respectively; (2) After each failed attempt to generate entanglement between Alice and Bob the electron is reset by optical spin-pumping~\cite{2013Natur.497...86B}. During this reset to $\mszero$ the nuclear spin can flip --- with $\Delta m_\mathrm I = \pm 1$ --- with a probability $p_\text{flip}$~\cite{2010Sci...329..542N}.

\begin{figure*}[htp]
    \centering
    \includegraphics[scale=1]{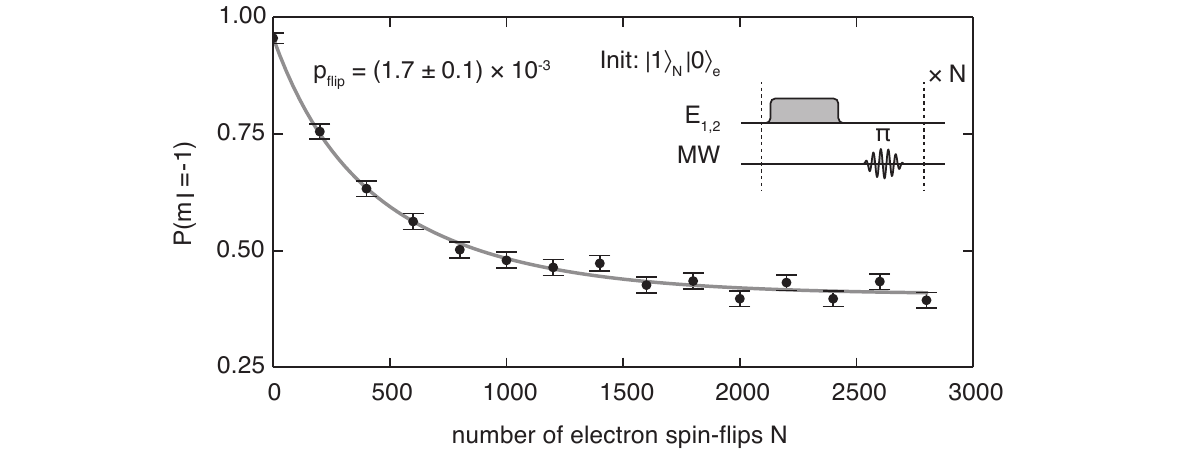}
    \caption{
    \label{fig:teleportation-SOM_Nspinflips}
    Nuclear spin state depolarization as function of electron spin flips by optical spin-pumping.
    We measure nuclear spin flips that are conditional on electron spin flips when optically pumping on $E_{1,2}$. We prepare the nuclear spin in $\msmone$ and measure the probability for its preservation dependent on the number of cycles of electron spin-pumping $\ket 1 \rightarrow \ket 0$ and re-preparation of $\ket 1$ by a microwave $\pi$-pulse. The solid line is a fit to the solution of (\ref{eq:teleportation-SOM_Nspinpopulations}) that is given by $p_{-1}(n) = 1/6 \left( 2 + (1-3p_\text{flip})^N + 3(1-p_\text{flip})^N \right)$ (neglecting initial population in $\mizero$ and $\mipone$). Because the data shown here is not corrected for finite initialisation fidelity of the nuclear spin and nuclear spin readout errors we include an offset $o$ and scaling factor $A$ in the fit function, $p_{-1}(n) = A/6 \left( 2 + (1-3p_\text{flip})^N + 3(1-p_\text{flip})^N \right) + o$. The fit yields a nuclear spin-flip probability of $p_\text{flip} = (0.17\pm0.01)\,\%$ per spin pumping cycle, and $A = 0.83 \pm 0.02$, $o = 0.13 \pm 0.01$. Note that the data shown in Fig.~2D of the main text has been corrected for nuclear spin readout errors. Error bars are 1 s.d.}
\end{figure*}

Assuming that the conditional probability for a nuclear spin flop accompanying an electron spin flip, $p_\text{flip}$, is identical for all $\Delta m_\mathrm I = \pm 1$, the equations describing the changes of populations in dependence of the number of electron spin flips, $n$, are
\begin{widetext}
\begin{align}
   \label{eq:teleportation-SOM_Nspinpopulations}
   p_{-1}(n) - p_{-1}(n-1) &= p_\text{flip} 
       \left( p_{0}(n-1) - p_{-1}(n-1) \right) \notag\\
   p_{0}(n) - p_{0}(n-1) &= p_\text{flip}
       \left( -2 p_{0}(n-1) + p_{-1}(n-1) + p_{+1}(n-1) \right) \notag \\
   p_{+1}(n) - p_{+1}(n-1) &= p_\text{flip}
       \left( p_0(n-1) - p_{+1} (n-1) \right).
\end{align}
\end{widetext}
The measured population of $\mimone$ in dependence of $n$ is shown in Fig.~\ref{fig:teleportation-SOM_Nspinflips}.

From independent calibration measurements we estimate the nuclear spin to be initialized by measurement with $p_{-1}(0) = 0.97$, $p_{0}(0) = 0.02$, and $p_{+1}(0) = 0.01$. Together with the nuclear spin depolarization during subsequent entanglement generation attempts we determine $\langle p_{-1} \rangle = 0.88$, $\langle p_{0} \rangle = 0.10$, and $\langle p_{+1} \rangle = 0.02$ from the solution of (\ref{eq:teleportation-SOM_Nspinpopulations}), for a maximum of 250 entanglement generation attempts before re-initialization by measurement. Here,
\be
    \langle p_{i} \rangle = \frac{1}{N} \sum_{n=0}^{N} p_{i}(n)
\ee
is the average population of the nuclear spin state $i$ for a maximum of 2N entanglement generation attempts. Note that the electron spin is in a superposition before reset, and thus the number of spin flips is half the number of entanglement generation attempts. The probability for successful entanglement generation is independent of the attempt number.

In the simulation of the experimental data we calculate the projected outcomes for each of the nuclear spin states and determine the combined result by weighing the average with $\langle p_{-1} \rangle$, $\langle p_{0} \rangle$, and $\langle p_{+1} \rangle$. Because population in $\mipone$ is outside the simulated space of two-level systems we treat this case in a separate simulation before weighing. The net effect of detuned MW pulses in this case is determined by calculating the electron spin rotation versus detuning and integrating over the $\mipone$ transition line shape.

The influence of imperfect nuclear spin initialization can also be approximated inituitively as follows: for $\mipone$ none of the operations on Alice's side are performed since all pulses applied are off-resonant, leading to dephasing of the state and ultimately a fully random outcome of Bob's readout. Initialization in $\mizero \equiv \ket0$ results in the opposite outcome than the one obtained from correct intialization in $\mimone \equiv \ket1$. Thus, with probability $2 \langle p_{0} \rangle + \langle p_{+1} \rangle$ the target state is fully mixed.

\subsubsection{Readout}
\label{sec:readout-error}

The major limitation of the Bell-state measurement fidelity is the finite single-shot readout fidelity of both electron and nuclear spin on Alice's side. Electron spin readout is achieved by resonant optical excitation of $E_y$. Detection of a least one photon during this interval is registered as readout result $\mszero$, otherwise the result is $\mspmone$. Nuclear spin readout is achieved by re-setting the electron spin to $\mszero$, mapping the nuclear spin state onto the electron spin by a CNOT, and reading out the electron spin. This procedure is performed twice in order to maximize the readout fidelity~\cite{Robledo:2011fs}. Readout result $\mizero$ is obtained for detection of at least one photon during either round.

The electron spin readout is limited by finite photon collection efficiency and electron spin mixing in the excited state~\cite{Robledo:2011fs}. For Alice, we measure a mean single-shot readout fidelity of $F_\text{e-RO} = 0.963 \pm 0.005$. The nuclear spin readout is additionally limited by the CNOT fidelity. With two readout rounds we estimate a mean readout fidelity of $F_\text{N-RO} = 0.985$ from the electron spin readout and CNOT pulse simulations.

In the simulation of the experimental results we use the single-shot readout fidelities to determine the conditional density matrices that arise after measuring the electronic and nuclear spin.

\subsection{Photon indistinguishability and entangled state fidelity}
\label{sec:lde-visibility}

The entangled state between the two electronic spins can be modeled as
\be
    \rho = V | \Psi^- \rangle \langle \Psi^- | + \frac{(1-V)}{2} \left( \ket{01}\bra{01} + \ket{10}\bra{10} \right ),
\ee
where the visibility $V$ describes the distinguishability between the photons emitted from Alice and Bob. Here we assume that all other imperfections are negligible compared to the photon distinguishability. The limitations of the Bell state fidelity are discussed in detail in Bernien {\em et al.}~\cite{2013Natur.497...86B}

For modelling we treat $V$ as a free parameter used to match the average teleportation fidelity. Using the parameters as described above and setting $V = 0.74$ (corresponding to a Bell-state fidelity of $F_{\Psi^-}$ = 0.87) our simulation yields a mean teleportation fidelity of $F = 0.77$.

\subsection{Further analysis of the teleporter performance}

\begin{figure*}[htp]
    \centering
    \includegraphics{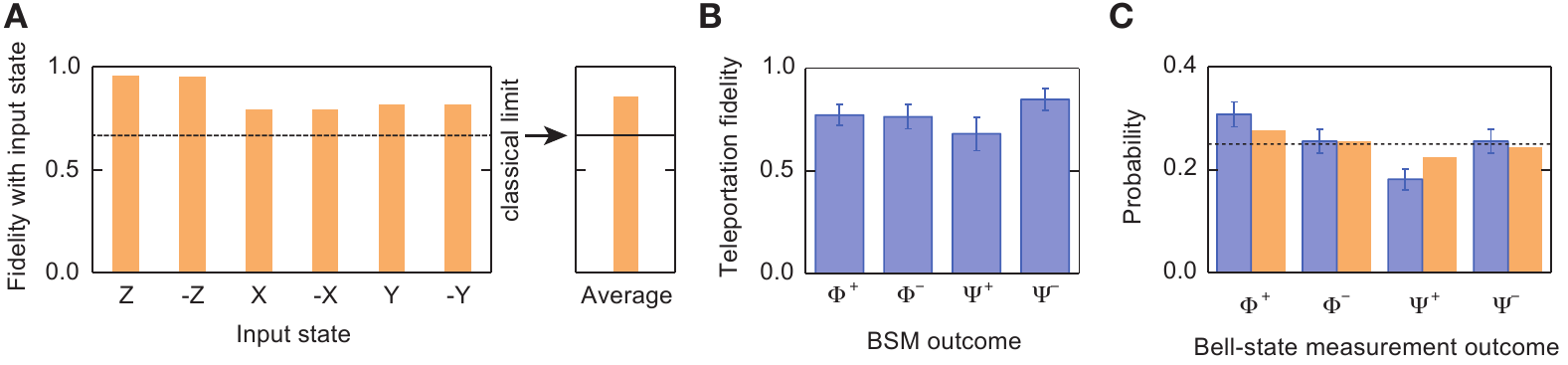}
    \caption{
    \label{fig:more_fidelities}
    Further analysis of the teleportation fidelity.
    \textbf{(A)} Correction for imperfect initialization of the source qubit. We simulate the teleportation outcomes using perfect intialization, $p_{-1} = 1$. The simulation yields and average fidelity of 0.86.
    \textbf{(B)} We determine the average teleportation fidelity for each outcome of the Bell-state measurement. Within the statistical uncertainty the fidelities do not differ substantially.
    \textbf{(C)} Probability for each BSM outcome, as measured (blue) and predicted from the model (orange). The dashed line marks 0.25. Error bars are 1 s.d.
    }
\end{figure*}

\subsubsection{Effect of the feed-forward operation}

Figure~3B of the main text shows the teleportation fidelity when no feed-forward is performed. This data is extracted from the teleportation data including feed-forward in the following way. We first determine the probability for obtaining the expected readout result independently for each BSM outcome by postselection. We then invert the readout result for all operations with a negative rotation sense. In this way we obtain the result that would have been measured if for each BSM outcome the same qubit rotation was performed (i.e., no feed-forward). We assume that any experimental errors in the final readout pulse are small and thus neglect them in this treatment.

\subsubsection{Correction for intialization}

After determining the entangled state fidelity as described above we can estimate the actual teleportation fidelity by assuming perfect intialization in our simulation. Setting $p_{-1} = 1$ we compute a mean teleportation fidelity of $F_\text{corrected} = 0.86$ (Fig.~\ref{fig:more_fidelities}A).

\subsubsection{Teleportation fidelity by Bell-state measurement outcome}

Due to the different readout fidelities for each of the four Bell states (see above and Fig.~2 in the main text) we can expect different teleportation fidelities as well. We find that the teleportation fidelity by outcome of the Bell-state measurement is consistent with expectations (Fig.~\ref{fig:more_fidelities}B), but the statistical uncertainty prevents a more detailed discussion.

\subsubsection{Probability of BSM outcomes}

We verify in more detail that the teleportation works as intended by examining the distribution of BSM outcomes obtained from all teleportation events (Fig.~\ref{fig:more_fidelities}C). The simulations are in good agreement with the data. The deviation from an equal probability of 0.25 for all BSM outcomes is mainly due to the asymmetry in the readout fidelities of the electron spin states \cite{Robledo:2011fs}.

\section{Acknowledgements}

We thank L.~Childress, L.~DiCarlo, M.~Hatridge, J.J.L.~Morton, A.~Reiserer, L.M.K.~Vandersypen, and P.~Walther for valuable discussions. We acknowledge support from the Dutch Organization for Fundamental Research on Matter (FOM), the DARPA QuASAR program, the EU DIAMANT and S3NANO programs, a Marie Curie Intra-European Fellowship and the European Research Council through a Starting Grant.


%

\end{document}